# Internet of Things Protection and Encryption: A Survey

Ghassan Samara
Computer Science Department
Zarqa University
Zarqa- Jordan
gsamarah@zu.edu.jo

Ruzayn Quaddoura
Computer Science Department
Zarqa University
Zarqa- Jordan
ruzayn@zu.edu.jo

Mooad Imad Al-Shalout
Computer Science Department
Zarqa University
Zarqa- Jordan
mooad_sh@yahoo.com

Khaled AL-Qawasmi
Department of Internet Technology
Zarqa University
Zarqa- Jordan
kqawasmi@zu.edu.jo

Ghadeer Al Besani
Computer Science Department
Zarqa University
Zarqa- Jordan
ghbesani@zu.edu.jo

*Abstract*— The Internet of Things (IoT) has enabled a wide range of sectors to interact effectively with their consumers in order to deliver seamless services and products. Despite the widespread availability of (IoT) devices and their Internet connectivity, they have a low level of information security integrity. A number of security methods were proposed and evaluated in our research, and comparisons were made in terms of energy and time in the encryption and decryption processes. A ratification procedure is also performed on the devices in the main manager, which is regarded as a full firewall for IoT devices. The suggested algorithm's success has been shown utilizing low-cost Adriano Uno and Raspberry Pi devices. Arduous Uno has been used to demonstrate the encryption process in low-energy devices using a variety of algorithms, including Enhanced Algorithm for Data Integrity and Authentication (EDAI) and raspberry, which serves as a safety manager in low-energy device molecules. A variety of enhanced algorithms used in conjunction with Blockchain software have also assured the security and integrity of the information. These findings and discussions are presented at the conclusion of the paper.

## I. INTRODUCTION

In today's digital era, the usage of the internet of things (IoT) in our lives is quickly expanding, for example, for practical and industrial applications. As more people utilize the internet of things (IoT) in research domains, (IoT) has attracted a lot of attention in order to give a simple and pleasant usage in everyday life [1, 2]. The Internet of Things (IoT) is described as a network that can process information by using sensors that are connected to the internet and can be directly linked to many devices over the internet. These devices are mainly reliant on their low cost. As a result, we may conclude that the actual growth factor of the Internet of Things is its low cost [3, 4, 5, 6]. Many people all around the world are interested in the (IoT) and are eager to gain access to sensor devices [7, 8]. It is worth noting that we require information security as well as data protection and that all devices connected to the internet should be more strict and powerful [9, 10]. The Internet of Things is linked to physical components or devices to provide real-time data to the Internet of Things data platforms, which will evaluate, comprehend, and offer real-time choices based on the rules created [11, 12]. Today's computer devices that are connected to the internet require assurances against the worries and breaches that may threaten data privacy, data hacking, and information security in the platforms and devices of the Internet (IoT) To offer a strong security communication method and a secure platform, it is necessary to organize information security controls and data encryption standards for each update cycle [13, 14, 15].

The usage of the internet is rapidly expanding, including smart homes, refrigerator temperature management, air conditioning, door locks, and other connected devices with sensors, regardless of whether they are permanently linked to the internet or receive data in real-time via sensors. According to estimates, the number of Internet-connected devices will reach 50 billion by the end of 2020/2021, and it is obvious that this number is rapidly rising [16, 17]. Given the popularity of the Internet of Things, a variety of suitable security measures in the internet of things devices are required to provide information security, data protection, and fraud detection. As a result, these devices may be utilized more effectively and safely [18, 19, 20]. One of the goals of this study is to protect privacy in all of its manifestations.

The rest of the paper contains the follows, the first section presents an introduction about the internet of things (IoT), the second section talks about the security threats that face the system of the internet of video things, the third section talks about IoT and blockchain encryption algorithms, the fourth section talks about improving the design of IoT algorithms, the next section talks about encryption process, the sixth section talks about evaluation and results, and the last section is the conclusion and future work. Then the references.

## II. SECURITY THREATS OF THE INTERNET OF THINGS

To fully integrate the industrial and societal sectors into the Internet of Things applications, many assurances and controls that improve information security and control privacy controls are required; otherwise, there will be a threat to data integrity due to easy access to physical objects that are not subject to supervision. This danger is regarded as the primary cause of security threats to Internet of Things applications, which are likely to affect the industry and the economy; it is also anticipated to raise worries about privacy breaches [21, 22, 23, 24].

The IoT devices are more vulnerable to security threats since their physical devices are tiny and are not subject to any of the previously stated supervisory mechanisms [25, 26]. As a result, controlling these devices as well as the external world surrounding them becomes challenging [27, 28]. Low-cost energy sources typically power IoT devices. This reason enables hackers to destroy any actual or tangible item connected to the Internet of Things via wireless media. An effective algorithm is developed to ensure that nothing from the device is leaked [29, 30].

It is possible that the algorithms currently in use do not provide the appropriate efficiency due to constraints imposed on computational resources and energy requirements because IoT devices have a small capacity while the energy used must be large to save the power of the algorithm as the large algorithm requires a large amount of energy that is greater than the energy available in IoT devices [31, 32].

The main structure of any IoT device includes two main tiers:

*1) Adequate data for the device is acquired when the sensor devices are connected.*

*2) The presentation tier: This tier is in charge of displaying the findings to the customer or Internet-connected devices.*

The components of the Internet of Things devices include a variety of optical sensors, devices for measuring temperature and humidity, devices for monitoring the proximity of sound and flow of chemicals, and other components that continually and actively monitor the environment [33, 34]. Although IoT devices have computing capability, they are incapable of processing huge amounts of data or activities. As a result, extra layers are required to deal with the data received from real devices [35].

Every IoT device has many tiers for sensing the needed data and delivering it from one device to another via one of the recorded tiers in a device-consistent manner. Extensible Markup Language (XML), JavaScript Object Notation (JSON), and separator formats are often used. The received information is encrypted utilizing Algorithms, which assure data confidentiality and protection from any compromise [36].

The most important function of data confidentiality in any system is linking consumer data in real-time [37]. Because the devices linked to the internet have tiny Peripheral devices, the data is encrypted and hidden from hackers to create a secure connecting channel. A comprehensive adaption is produced for sharing data to maintain privacy. They might go unmonitored for days or weeks [38]. This situation puts them at risk of data theft and assaults. As a consequence, hackers can access the device's main memory and extract data. To avoid this, it is critical to use robust data encryption and expand the encryption keys' work to become extremely secure and cannot be readily recovered [36].

This provides a starting point for protecting Internet of Things devices from network threats and ensuring data integrity. Many suggested algorithms, such as Advanced Encryption Standard (AES), LEA, and RC6, have been developed and deployed in IoT devices. However, because IoT devices are limited to a particular and basic power, less processing power is acquired. Thus we are striving to develop a large algorithm with few cycles [39, 40]. The communication channel between IoT devices and IoT platforms should be secure and unaffected by any threat [41].

Because of the growth in threats on the internet of video things, which exceeded two thirds, there is a constant need to enhance the algorithm and examine the data. According to the Sonic Wall study for IoT attacks on the internet over 2019-2020 at rates surpassing 66 percent. In 2020 [42, 43], the number of attacks grew substantially from 34.3 to 56.9 million attacks. Attacks on the Internet are often classified as a binary attack in the middle of a physical attack, seizing the device's memory, or breaking the Internet connection lines [44, 45]. Data encryption aids in the transport and conversion of plain texts into encrypted texts, therefore minimizing the potential of data theft. In any case, there is a considerable reduction in the performance of the present algorithms, and they cannot fulfil the requirements of low-power IoT gadgets [46]. As a result, improved techniques for protecting tools and devices against known threats and data from hackers must be developed [47, 48].

Many algorithms for the Internet of Things have been proposed. A new structure was also proposed to connect IoT devices using high computing and blockchain technology, which is an improved algorithm that depends on the speed of encryption and the connection to the headquarters that is responsible for protecting the internet of things from attacks and illegal entries by breaking data into blocks using the Blockchain algorithm. In this work, we explore improved algorithms for IoT devices that are less powerful and have weak computing capabilities. These features allow us to contact the building's security management and request that IoT devices and tools be installed. A new structure is also offered to introduce the Blockchain in Node Manager to handle data integrity and IoT device integrity inside the network [49].

## III. IOT AND BLOCKCHAIN ENCRYPTION ALGORITHMS

According to the sufficient explanation, it is vital to spot the shortcomings of the algorithms in various low-powered devices with limited resources. As a result, if any network node is targeted [50] and then controlled, the entire internet of things becomes vulnerable to a wide range of other attacks. To that aim, we propose considering blocked zeros as viable possibilities for activating restricted devices via secure media. According to the block chain algorithm, crypto-analysis assaults, such as linear and differential techniques, are one of the difficult ways in Feistel encryption. They can use S-Box functional correlation layers, confusion, and 4-bit propagation. All characteristics are defined by the plain text of a 64-bit long entry, the 48-bit encryption key, and the block size of 64 bits, 80 bits, and 128 bits. Long keys are what they're called. When the round key is pushed, the preset algorithm begins the process of encoding substitution or flipping blocks. Because the Feistel structure allows for various device groupings, as advocated by the National Security Inspectorate (NSI), Simon's technique is the best device implementation method. The letters a and A2 are widely used to signify block size and key length, respectively. The SIMON algorithm allows for a maximum block size of 128 bits. It is a strong alternative to (AES) encryption since it is versatile enough to be used in any IoT board and device without interfering with the devices.

The (LILLIPUT) [51] technique employs improved block encryption with a block size of 64 bits and an 80-bit key length. As a result, it is created in an S-Box style similar to Present. [52] The SB network is the basis for KLIN block encryption. It provides 64-bit blocks. The length of the key can range between 64 and 96 bits. Both messages and hash values are authenticated. The exchange value of this key is not static, and the keys were previously switched dynamically with a set of specified values. RC5, Tiny Encryption Algorithm (TEA), and other algorithms have been proved to be effective in Block Cipher (XXTEA) [53]. These algorithms require higher processing power. As a result, increasing the cycles and regularly changing the key for the specified device can result in the initial key recovery. A survey was undertaken to discover that various algorithms give an effective and secure method of protecting plain text while linking the particles of the internet of things. However, when it comes to data integrity, the algorithms fall short of providing the essential safeguards. [54] As a result, it is suggested that a smooth, lightweight sequence of the blocks used in the IoT environment be implemented to protect the integrity of the data provided in each portion and every new connection on IoT platforms. [55] A group proposed a research paper in which the particles of the Internet of Things were organized into a special network in which the high computing device would work within the node, provide authentication and direct packages if there are a large number of movements within the network. It could be decoded in an indexed form in the blockchain within the hardware of smart devices already present at homes. A researcher named LIE [56] described the structure of the blockchain feature for vehicle communication. He added that whenever the vehicle moves to a new zone, the heterogeneous blockchain group receives a detailed message about the vehicle and its main information. In addition to the major information about these cars, this subject provides additional information about them for organizers or communities interested in tracking vehicles and collecting data about them by recognizing the movement of these vehicles in real-time. This understanding reduces the need for encryption when sending and receiving data over mobile phone networks. It also takes into account current security concerns concerning traditional data transfer protocols.

IV. IMPROVING THE DESIGN OF IOT ALGORITHMS:

Table 1 shows how the enhanced (EDAI) algorithm is used for authentication and data integrity for IoT devices using widely used substitution-permutation networks (SPNs) and the Festal encoding method [57]. This method functions dynamically, merging the features of the Language Server Protocol (LSP) and VESTL into a single form to boost the security of IoT particles in fewer encryption rounds than currently available techniques. The encryption rounds range from 20 to 64 [58], depending on the key size. Because IoT particles have limited energy and resources, reducing encryption frequency to a bare minimum and establishing more complicated keys ensures better production and extraction of a coded text in a regulated environment. The following is a list of IoT devices that are regularly used.

TABLE I. CATEGORIES BLOCK CIPHER [31]

| SP Network | Festal Architecture |
|---|---|
| AES | DES |
| 3Way | Blowfish |
| PRESET | SF |
| SHARK | Camellia |
| SAFER | LWV |

The use of the Feistel cypher via the SP network and the AES method has the advantage of reducing the complications that develop during the encryption and decryption operations on distinct works. [59] From another perspective, there is a related technique for generating cypher text (CT) and then converting it back to plain text (PT). The (EDIA) technique encodes a defined segment of asymmetric 64-bit key particles, where each key and text are the same size and encoded an agreed-upon number of times, denoted by n. Encryption is achieved by combining keys and bits in order to achieve optimum results. It is recommended that five rounds of encryption be used with a 16-bit key length instead of the original key's 64 bits. To stop the attacks in the main search phase, a number of repetitions must be created to make the used key more difficult [60].

TABLE II. KEY CONFIGURATION [60]

| KEY – Configuration | Example |
|---|---|
| Firmware ID | ADZ32xcfrcv |
| Current Data Timestamp | 202020 12 : 12 : 10 |
| Version ID | 2.2 |
| Checksum | A1 (A-z, 0-9) |

This sequential format is produced based on a prior sequence. Take a look at the data in Table 2. The checksum logic is determined for each Internet-of-Things object or tool as a result of the checksum calculated from the current timestamp and firmware ID in such a way that only the sender and the receiver maintain it. If any communication error occurs during optional assembler creation of the key or timestamp, it will lead to a mismatch in the algorithm's key.

Key config =c ( K ( firmware ID , Time , Version ) , Fun key )
.......... (1)

c → check sum Function
K→ Key Generation Function .
Fun c key → Encryption – E \\ Decryption – D

The 64-bit keys' outputs are not simple text or a default key provided by the Internet of Things device, but rather a series of functions that allow higher complexity to generate and expand the key. To ensure the key's complexity, we work on four primary values with special logic to build a two-bit audit. [61].

*Master schedule - time expansion:*

The EDIA algorithm is a key generation algorithm that is used in mathematical functions. It generates four circular keys of 16 bits each. Each key is utilized in cypher rounds. It can also be used with a subkey. The substitution and switching approach is used while entering the plain text in order to receive the encrypted text at the end of the encryption process. In most cases, the tour is solely accountable for everything, like as hacking or transmitting unencrypted text.

The method used is the one that agrees on the primary inputs from the source text sector, and it includes the Bit Randomize property:

It is the one who creates the master key and ensures the key's complexity while also preventing attackers from obtaining the primary key or subkey, as shown in fig 1 [61].

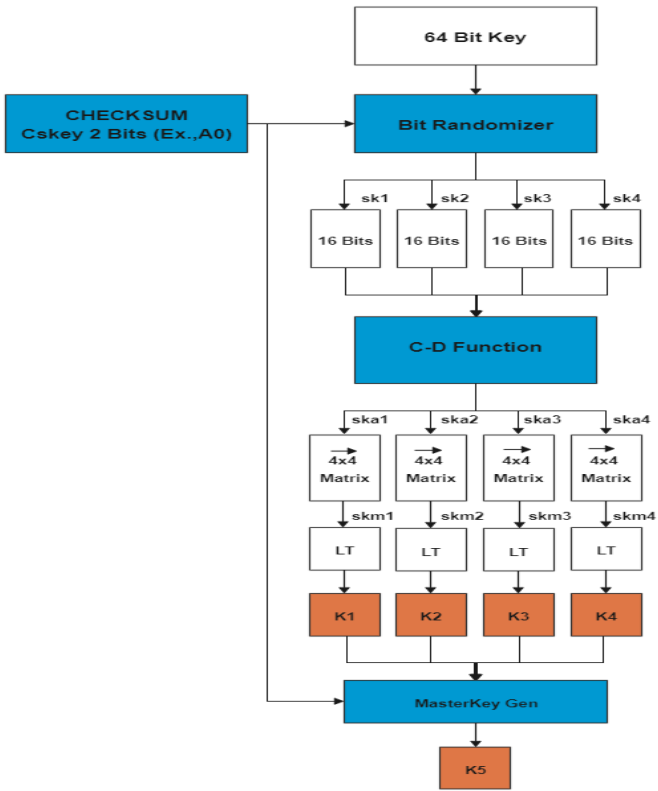

Fig. 1. Key Expansions Process [61]

The main objective of Bit Randomize is to assign bits at random from the master key to its subkeys (6 * 14 bit) (Sk1, Sk2, Sk3, Sk4). Each key is a four-by-four bit combination of the generated key. Based on the first piece of information, this is the case. Checksum As illustrated in table 3, a 64-bit key is divided into a 16-bit intermediate key.

TABLE III.    BIT RANDOMIZER METHOD [62]

| Bit Randomizer – Key text 1st Randomizer | |
|---|---|
| 1 | Input 64 Bits Key (KP) Using fire ware, Timestamp, Checksum |
| 2 | Output 4*16 Bits sup key |
| 3 | P Array [16]=Key |
| 4 | Get the First bit of check sum ( 1: 1 ) |
| 5 | IF c sum = = Alphabet sequence → Randomizer |
| 6 | A-G → Randomizer ( P Array [ 1,5,9,13] ) |
| 7 | H-N → Randomizer ( P Array [ 16,12,8,4] ) |
| 8 | O-T → Randomizer ( P Array [ 2,6,10,14] ) |
| 9 | U-Z → Randomizer ( P Array [ 3,7,11,15] ) |
| 10 | Repeat Step 6 to 9 |
| 11 | END |

As shown in Table 4, the C-D function is a function that uses jamming and propagation to switch bits using tables. These tables combine the predefined bits, which are employed in two more tables, C and D.

TABLE IV.    C-D FUNCTION TRANSFORMATION [63]

| K c – Cipher Key | C Transformation | D Transformation |
|---|---|---|
| 0 | E | 6 |
| 1 | 3 | 1 |
| 2 | A | B |
| 3 | 9 | 5 |
| 4 | F | 0 |
| 5 | 2 | E |
| 6 | C | 3 |
| 7 | 8 | A |
| 8 | D | 9 |
| 9 | 7 | F |
| A | 4 | 4 |
| B | 6 | 2 |
| C | 1 | C |
| D | B | 8 |
| E | 5 | D |
| F | 0 | 7 |

Normally, the C - D function produces four 4 * 4 matrices. Skm1–Skm2–Skm3–Skm4–Skm5–Skm6–Skm7–Skm8–Skm9 The matrices are based on the C and D conversion and are supplemented with bit combinations that result in four keys (Kr) [63].

Creating Master Keys: It collects 4 * Kr and performs an XOR operation on its output. It includes the second half of the total that was tested and displayed when the first key was generated. Figure 2 shows how to add a third master key to complicate the encryption and decryption operations further.

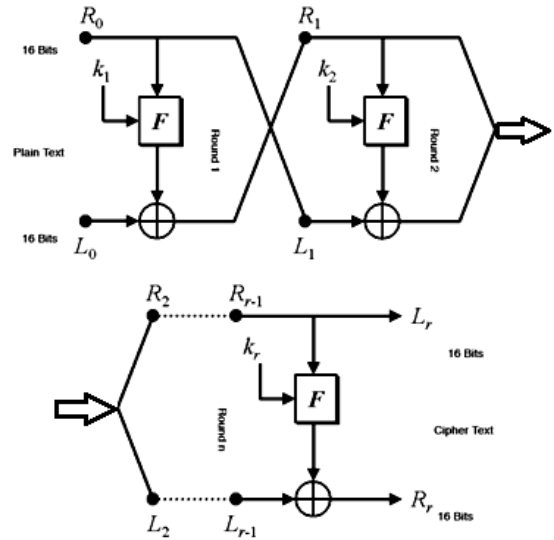

Fig. 2. Encryption With XOR Process [36]

## V. ENCRYPTION PROCESS

The initial procedure Feistel uses the same processes and rounds for encryption and decryption, as shown in the diagram below, and it works with 16-bit data in four blocks for a total of 64 bits. Each of the 16 blocks of plain text (Pt) and cypher text is encrypted at the end of each round (Ct). The (EDIA) technique uses a five-point encryption strategy, with each 16-bit combination serving as an input and the bits being changed regularly to increase the security and complexity of the coding text. K1-5 is an expansion product that is frequently used in any round. In addition to the XNOR technique, C-D and the diffusion function are employed to overlap texts if they are available.

The C-D function is used to transfer L0 and R0, after which the C and D transformations are applied, and the final values are as follows:

$$Ct = Lr1 + Rr1 + Lr2 + Rr2 \ldots\ldots\ldots (2)$$

The process for exchanging keys both enhances and complicates the cipher script. Furthermore, as previously stated, the master key function works on the embedding process via checksum as an additional security component to reduce attacks on the master texts.

- A simple blockchain for the Internet of Things:

The blockchain uses records to protect data distribution and storage. A complex computing process performs the combination by distributing the data. As a result, attackers cannot obtain any form of data or hack the data available on any network because all of the remaining nodes will not respond to the responses, resulting in inconsistent data, as shown in fig 3 [36].

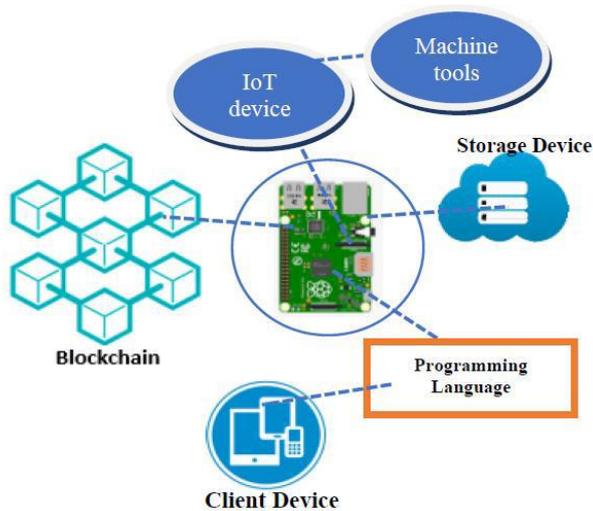

Fig. 3. Block chin Implementation [63]

TABLE V.  AUTHENTICATOR LOGICS [64]

| | IOT Node Authenticator |
|---|---|
| 1 | Input : Request to add the data and resources to Block |
| 2 | Output : Pass- fail |
| 3 | Validate the check sum and Node physical ID |
| 4 | Solve the Node _ problem () using the check sum |
| 5 | If Node _Problem () == " Pass " |
| 6 | Access to Requested Resources |
| 7 | Else |
| 8 | Add the Node to Review    List () |
| 9 | Send the Dummy Simulated Response () |
| 10 | Cascade the Info to all Node Mangers |
| 11 | Increases the Risk ID of adjacent Nodes |
| 12 | END – IF |
| 13 | Respond to the Node Request |
| 14 | Initiate the Validate Review –List Task |

The authentication of the Internet of Things is a sign and one of the most important acts stated in the authentication tool sequence. The table below represents a complex and difficult-to-solve authentication challenge. Because all particles are semi-connected and in private networks, direct connections to public networks are limited in a simple approach that assures data and particle integrity. Despite increased pressure, data protection and privacy can be provided by using an additional device, regardless of material costs [64].

If the code to be encrypted is: 8787878787878787  And my DES Key specified is: 0E329232EA6D0D73, we can get the ciphertext is: 0000000000000000. if the same DES key were used for the above cypher text 0E329232EA6D0D73, the original text 8787878787878787 would be the result of the decoding.

Examples are employed in this scenario because the plain text is 64 bits long, and the blain text can also be 64 bits long. However, in most circumstances, plain text is not a 64-bit (16-hexadecimal digit) integer multiplication.

This language computer code was used in the context of encryption using the DES and AES algorithms.

VI. EVALUATION AND RESULT

Many important elements must be considered while evaluating the (EDIA) algorithm in (IoT) devices, such as resource use, implementation time, maximum energy utilization, and key complexity. Aside from the encryption and decryption operations, the sensitivity of the key to the input process is controlled by a sufficient number of employed rounds. To examine all data, a comparison is done between the algorithm used and the previously reviewed algorithms and gives the most accurate results. From this vantage point, we work our way through the memory available for the method to do various arithmetic operations on plain text. If the memory usage is high, the Internet of Things device will be unable to keep the memory as the number of rounds increases. As a result, because memory is one of the most important things, we must use it optimally. Any IoT device's power is limited in the middle of the rounds, as is the time it takes to generate plain text for encoding and decoding.

Furthermore, it restricts other functions of devices with limited energy. Here is where the assessment algorithm comes into play for the required suitability in terms of energy and required functionalities. Provides a method of attack the search key is the attacker's ability to obtain the key and change the encryption text. The sensitivity to the fake key must typically be high enough that retrievers are unable to recover the original content. The algorithm's performance is often measured using the AT Mega 328 Adriano board software, which attempts to conserve energy while accepting data from an external source. The time spent implementing the encryption is typically 0.195 decoding is 0.190 milliseconds. The algorithm's memory, which is 30 bytes, is used to generate texts for people who want to encode them, and the comparisons with algorithms are presented below:

TABLE VI.   TABLE 6. : COMPARISON OF NUMBER CYCLE  [63-64]

| Algorithm | Hardware | Block | Key length | Lines | Memory | Encryption of NO. cycle | Decryption of NO. cycle |
|---|---|---|---|---|---|---|---|
| KATAN | DVR | 64 | 80 | 338 | 18 | 72063 | 88525 |
| KLEIN | DVR | 64 | 80 | 1268 | 18 | 6095 | 7658 |
| EDIA | ATmega328 | 64 | 64 | 950 | 30 | 3100 | 3048 |

The energy consumption is computed based on a number of criteria, including the number of encoding and decoding cycles, and it is close to 160.24 micro joules. Total transmission cycles have roughly 1800 micro joules. As the complexities of the encryption and decryption key are regularly enhanced, the information appears to be the better usage that can be used for any device, IoT device, or tool.

## VII. CONCLUSION AND FUTURE WORK

This work presented and provided integration between data and security via a lightweight (SPN) and Feistel algorithms combined through 5 rounds of encryption and decryption procedures on Adriano systems. Security analyses yielded positive results. Raspberry Pi Nodal Manager devices safeguard Adriano particles with limited energy and resources. They ensure particle integration in IoT networks, which consume the speed of encryption and data block ratification. We achieved good results by reducing the number of attacks and modifying particles and their penetration with lightweight algorithms and virtually during confirmation.

We can improve internet of things (IoT) devices by increasing access to random memory (ROM) and random access memory (RAM) in many used devices with limited resources and applying the same things to high-end devices in terms of energy to assess the feasibility of using such an approach in regular devices in homes, smart factories, and smart logistics.